# Phase-Incremented, Steady-State Solution NMR: Maximizing Spectral Sensitivity Without Compromising Resolution

Mark Shif,[1] Yuval Zur[2,*], Adonis Lupulescu,[1] Tian He,[1,3] Elton T. Montrazi,[1] and Lucio Frydman[1,*]

[1]Chemical and Biological Physics Department, Weizmann Institute, Rehovot, [2]Insightec Ltd, Tirat Carmel, Israel; [3]Chemistry Department, Zhejiang University, P. R. China

*Corresponding authors: yuvalzur50@gmail.com; lucio.frydman@weizmann.ac.il

## Abstract

NMR acquisitions based on Ernst-angle excitations are widely used in analytical spectroscopy, as for over half a century they have been considered the optimal way for maximizing spectral sensitivity without compromising bandwidth or peak resolution. However, if as often happens in liquid state NMR relaxation times $T_1$, $T_2$ are long and similar, steady-state free-precession (SSFP) experiments can actually provide higher signal-to-noise ratios per $\sqrt{\text{acquisition\_time}}$ ($SNR_t$) than Ernst-angle-based counterparts. Although a strong offset dependence and a requirement for pulsing at repetition times $TR \ll T_2$ leading to poor spectral resolution have impeded widespread analytical applications of SSFP, phase-incremented (PI) SSFP schemes could overcome these drawbacks. The present study explores if, when and how, can this approach to high resolution NMR improve $SNR_t$ over the performance afforded by Ernst-angle-based FT acquisitions. It is found that PI-SSFP can indeed often provide a superior $SNR_t$ than FT-NMR, but that achieving this requires implementing the acquisitions using relatively large flip angles. As also explained, however, this can restrict PI-SSFP's spectral resolution, and lead to distorted line shapes. To deal with this problem we introduce here a new outlook on SSFP experiments that can overcome this dichotomy, and lead to high spectral resolution even when utilizing relatively the large flip angles that provide optimal sensitivity. This new outlook also leads to a processing pipeline for PI-SSFP acquisitions, which is here introduced and exemplified. The enhanced $SNR_t$ that the ensuing method can provide over FT-based NMR counterparts collected under Ernst-angle excitation conditions, is examined with a series of $^{13}C$ and $^{15}N$ natural abundance investigations on organic compounds.

# 1. Introduction

Nuclear Magnetic Resonance (NMR) is an essential tool in contemporary chemistry, widely used in both academia and industry to derive molecular structures, dynamics and concentrations.[1-5] Whether carried on liquid or solid samples, whether focusing on glasses or on living organisms, nearly all NMR experiments follow Anderson-and-Ernst's Nobel-award-winning Fourier Transform (FT) proposition.[1-7] That is not surprising as FT-NMR is simple, general, and provides excellent resolution while covering large spectral bandwidths. Furthermore, and crucial in its eventual adoption as *the* way of collecting NMR data, FT-NMR exhibits better signal-to-noise ratio per $\sqrt{\text{acquisition\_time}}$ ($SNR_t$) than alternative frequency-domain-based approaches. However, it is also known that if spectral resolution and the faithful coverage of peak intensities over large bandwidths are not a must, the Anderson-Ernst FT-NMR proposal is not necessarily optimal for achieving the highest $SNR_t$: when $T_1 \approx T_2$, as is often in solution-state experiments, and if a peak's offset and the excitation pulse angle can be chosen at will, Carr's steady-state free-precession (SSFP) NMR can often provide a superior $SNR_t$.[8-10] While instances where SSFP could become a method of choice in spectroscopic applications have been described in low-resolution and in solid-state NMR and NQR,[11-13] the aforementioned limitations have constrained SSFP's use to MRI –where it is exploited under variety of vendor-dependent acronyms.[14-17] $SNR_t$ advantages were also here *the* defining reasons of why MRI, a widely used yet costly medical imaging modality where both scanning duration and data quality are of essence, was quick to adopt SSFP. Indeed, with a focus on a single water resonance whose offset can be chosen more-or-less at will and with $T_1$ and $T_2$ times that are reasonably close, MRI was also uniquely posed to deal with SSFP's main two drawbacks: its lack of spectral resolution, and its strong dependence on the offset (i.e., the chemical shift) of the targeted peaks. SSFP's spectral resolution limitations can be adumbrated from its pulse sequence, which involves a train of closely-spaced pulses with constant flip-angle α, applied at repetition times $TR \ll T_2, T_1$ (Fig. 1, top). This leads to signals S(t) that are a combination of free induction decays (FIDs) and of multiple echoes –the latter reflecting, at each TR, a sum of histories associated to different coherence transfer pathways[1-4] (or as known in the MRI literature, to different extended phase diagrams[18,19]) refocusing at the top of every pulse in the sequence. SSFP's $SNR_t$ potential and offset dependence –qualities which will both be central in this study– are highlighted in Figure 1 (bottom). It follows from these extensively verified predictions that steady-state pulses can lead to transverse magnetizations reaching up to 50% of the thermal equilibrium magnetization, in a nearly constant emission of NMR signals. Such feat, however, requires the resonance being addressed to have a suitable, *a priori* known offset Δ, and the use of relatively large flip angles α. The issue of offset-dependence has been particularly detrimental in high-resolution NMR settings: given SSFP's repeated pulsing, offsets will arise (e.g., an on-resonance situation) where a steady, large-flip-angle pulsing with a constant phase (e.g., *x*), will lead to a null ($M_{x,y} \approx 0$) signal. Furthermore, given SSFP's periodicity, its unusual, offset-dependent excitation profile will repeat itself modulo $2\pi/TR$. When coupling to this alternating dark/bright spectral pattern the limited spectral resolution arising from its demand for short TRs, it is clear why SSFP's $SNR_t$ advantages were no match against the generality and convenience of FT-NMR. Thus SSFP, together with related driven-equilibrium options,[25] have remained in the fringes of spectroscopic NMR applications.

Very recently, driven both by curiosity and by promises of increased $SNR_t$, we revisited SSFP's potential in a number of high resolution spectroscopic applications.[20,21] Arising from these spectroscopic studies was the realization that some of SSFP's main weaknesses –strong offset sensitivity, periodic regions of high and null intensities, repeated folding patterns– may actually contain the seeds for achieving high resolution among inequivalent sites over large bandwidths, as demanded by analytical liquid-state NMR.

Based on this we recently proposed a novel approach to attain high resolution NMR information from SSFP signals, based on monitoring the steady-state responses over a series of offsets $\{\delta_m\}_{0 \leq m \leq M-1}$ covering the $1/TR$ folding interval, and then exploiting the *a priori* known dependence of a peak's intensity on offset to pinpoint the latter within such interval.[22] In addition, the extreme fold-over associated with SSFP acquisitions was dealt with by a discrete FT of the short FIDs that were sampled within the inter-pulse intervals *TR*. Solution-state $^{13}$C NMR spectra which compared well with FT-NMR data could then be obtained on simple organic compounds, using this phase-incremented (PI) SSFP approach. Still, given that line shapes in these PI-SSFP spectra did not arise from a FT, their sensitivity and point-spread functions (PSFs) possessed a number of distinctive properties. Most remarkable among them was a dependence of the spectral resolution on the flip-angle α used in the excitation of the spins; this is in marked contrast to resolution in FT-NMR, where line widths are dictated by the duration of the acquired free-induction decay (FID) signals. Indeed, in PI-SSFP NMR arbitrary short acquisition times are no impediment for resolving closely-spaced resonances, as it is the combination of α flip angles being applied and of offsets *M* being interrogated –but not necessarily to the FID acquisition time given by *TR*– that govern PI-SSFP's ability to distinguish inequivalent sites. The fact that relatively small αs were then needed to obtain the customary, Hz-sized resolution was then a mixed blessing: on one hand it freed peak intensities from the usual $T_1$-weighting that affects FT-NMR spectra, but on the other it prevented the maximization of the SSFP $SNR_t$ potential –which as illustrated by the $|M_{x,y}|$ values presented in Fig. 1, benefits from larger flip angles when considered over an arbitrary range of offsets.

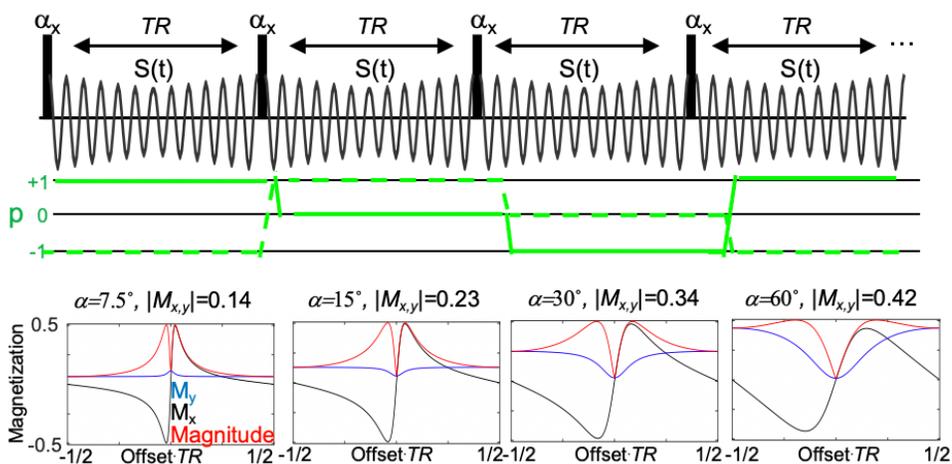

**Figure 1.** (Top) SSFP sequence involving a train of pulses α (here applied with constant phase) spaced by time intervals TR. Shown underneath is a subset of the coherence transfer pathways undergone by an isolated spin ensemble under the action of the sequence, illustrating the complex echoing at the top of any given pulse. (Bottom) Steady-state responses S(0) arising at the top of each SSFP pulse/echo as a function of a site's frequency offset, shown for four different flip angles α. Shown as well are the transverse magnetizations arising for each flip angle over the full frequency range, and their integrated absolute values. Offsets past ±1/2TR repeat themselves inside this region by repetitive fold-overs.

The present study revisits the origins of PI-SSFP's demand for small flip angles in order to achieve high spectral resolution. It is shown that it is not because of fundamental principles but rather because of instabilities in the data processing, that the resolution of inequivalent peaks in PI-SSFP NMR is complicated when relying on large α flip angles. It is also shown that for common instances, particularly when dealing with sites possessing relatively long longitudinal relaxation times $T_1$, overcoming these instabilities could lead to substantial $SNR_t$ advantages over the FT-NMR scheme which has served as backbone of analytical spectroscopy for the last half century. A generic PI-SSFP formalism coupled to a tailored processing algorithm that, based on such derivations, can deal with large α-angle instabilities, is then put forward. The result is a new approach that can provide a spectral sensitivity that matches or exceeds that of 1D FT-NMR experiments based on Ernst-angle excitations, at no compromise in spectral resolution. This is exemplified with a variety of 1D $^{13}$C and $^{15}$N NMR data. Limitations as well as additional potential extensions and applications of this novel approach to high-resolution NMR, are briefly discussed.

## 2. Theoretical Background

**2.1 A deterministic approach to the 1D PI-SSFP NMR spectral reconstruction**. As mentioned, the PI-SSFP proposal to collect high resolution 1D NMR spectra relies on acquiring and processing an array of (signal-averaged) steady-state FIDs of duration *TR*, as a function of offset. This is best carried out by signal averaging a series of NS scans, where the carrier offset is constant but the phases of consecutive RF pulses are incremented in steps of $\{\varphi_m = 2\pi\, m/M\}_{0 \leq m \leq M-1}$ (Figure 2a). This will provide an array of *M* FIDs $\{S_m(t)\}_{0 \leq m \leq M-1}$, sampled over times *0<t<TR*. Assuming for simplicity the presence of a single peak positioned at a frequency *f*, and given the aforementioned $1/TR$ periodicity of the SSFP signal response *S* on *f*, it is possible to describe the ensuing *M* steady-state FIDs at times t=0 (e.g., immediately after each pulse) as the discrete Fourier series[23,24]

$$S_m(0) = S(0, 2\pi f \cdot TR + \varphi_m) = I(f) \cdot \sum_{k=-\infty}^{\infty} A_k \exp(ik\varphi_m) \cdot \exp(ik\, 2\pi f\, TR)\,, \quad m = 0\ldots M-1. \quad [1]$$

where $I(f)$ is the spectral intensity of the peak at frequency *f*. The $A_k$ in Eq. [1] are discrete Fourier coefficients reflecting the multiple transfer pathways depicted in Figure 1, that can be calculated analytically and whose amplitude decays to zero as |k| increases.[23,24] These coefficients depend on the flip-angle α, on E$_2$ = exp(-TR/T$_2$) and, more weakly, on E$_1$= exp(-TR/T$_1$); importantly, *the larger the flip-angle α or the shorter the T$_2$, the faster the $A_k$ will decay to zero with |k|*. We have recently shown[22] that knowledge of these $A_k$ coefficients –whether from analytical or numerical sources– can allow one to find from the $\{S_m(0)\}_{0 \leq m \leq M-1}$ set, the spectral amplitude $I(f)$ that, after multiple potential foldings within the ±1/2TR interval, will characterize a peak of frequency *f*. Assuming for the sake of simplicity that the frequency being searched for falls within such interval, i.e., that $-\frac{1}{2\text{TR}} \leq f \leq +\frac{1}{2\text{TR}}$ (and that therefore fold-overs can be disregarded), we proposed recreating the NMR spectrum within such range by using a linear combination of the $S_m(0)$ signals. Denoting this linear combination as *F(f)* and the *a priori* unknown coefficients that will be involved in it as $\{\beta_m\}_{0 \leq m \leq M-1}$, we can define the linear combination as

$$F(f) \triangleq \sum_{m=1}^{M} \beta_m S_m(0) = \sum_{k=-\infty}^{\infty} \sum_{m=1}^{M} A_k \exp(ik\varphi_m) \cdot \beta_m \cdot exp(ik\, 2\pi f\, TR) \quad [2]$$

Achieving high spectral selectivity means that this linear combination function should mimic as closely as possible a discrete bandpass filter, whose line shape will define the "peaks". Based on filter response theory, this filter can be written as

$$R(f) = \sum_{k=-N/2}^{\frac{N}{2}-1} C_k \cdot \exp(ik\, 2\pi f\, TR) \quad [3]$$

where the $\{C_k\}_{-N/2 \leq k \leq N/2-1}$ coefficients can be calculated based on a desired response (e.g., using the Finite Impulse Response script in Matlab's signal processing toolbox[25]). Finding the linear combination that makes *F(f)* as close as possible to *R(f)* then demands solving a series of linear equations

$$C_k \approx A_k \sum_{m=0}^{M-1} \beta_m \cdot \exp(ik\varphi_m)\,, \quad k = -\frac{N}{2}\ldots\frac{N}{2}-1 \quad [4]$$

Suitable solutions $\{\beta_m\}_{0 \leq m \leq M-1}$ of these equations will generate a narrow filters with a targeted width of $1/(TR \cdot NB)$ –NB being the total number of bands (peaks) to be resolved within $\pm 1/2TR$. Eq. [4] can also be written in matrix form as

$$\mathbf{C} \approx \mathbf{L} \cdot \boldsymbol{\beta} \qquad [5a]$$

where $\mathbf{C}$ is a N-by-1 vector containing the $\{C_k\}_{-N/2 \leq k \leq N/2-1}$ coefficients, $\boldsymbol{\beta}$ is an M-by-1 vector with the $\{\beta_m\}_{0 \leq m \leq M-1}$, and $\mathbf{L}$ is a N-by-M matrix whose $k,m$-element is

$$L_{k,m} = A_k \exp(ik\varphi_m) \quad . \qquad [5b]$$

The $\beta$-coefficients needed to define the intensity of a peak at frequency $f$ could then be found by minimizing the norm $\| \mathbf{C} - \mathbf{L}\boldsymbol{\beta} \|_2^2$ using the Moore-Penrose inverse matrix $\boldsymbol{\beta} = (\mathbf{L}^\dagger \mathbf{L})^{-1} \mathbf{L}^\dagger \mathbf{C}$, where $\mathbf{L}^\dagger$ is $\mathbf{L}$'s conjugate transpose.

Figure 2 clarifies further these arguments, by describing how this proposal to high resolution NMR (sequence in Fig. 2a) finds a "peak intensity" at $f = 0$. Illustrated in Fig. 2b is the SSFP response vs phase increment for this site at zero chemical shift, for the array of phase-increments used by the sequence and as a function of increasing flip angles $\alpha$. Notice that the overall signal magnitude increases steadily with $\alpha$, but –as adumbrated by the plots in Fig. 1– the offset (i.e., the $\varphi_m$) dependence of the SSFP response also "flattens", and thereby loses frequency discrimination insight. This is reflected by the $\{A_k\}$ coefficients

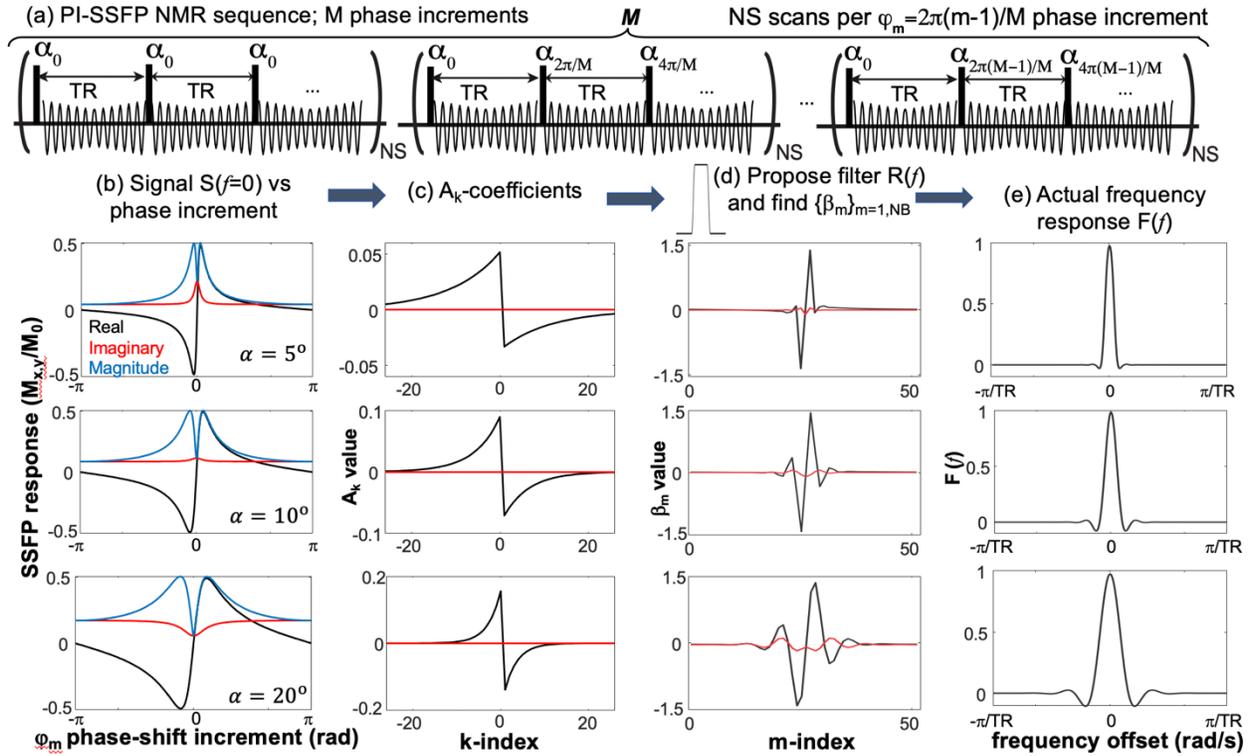

**Figure 2.** (a) PI-SSFP approach to high-resolution NMR, involving a train of NS signal-averaged scans excited by pulses of flip-angle $\alpha$ spaced by a time TR, and relative phases $\varphi_m$ incremented as shown over M uninterrupted experiments. (b) Single-site SSFP response vs relative phase increment $\varphi_m$ for different flip angles $\alpha$, assuming $f=0$, $T_1=5$s, $T_2=2$s, and constant (zero) receiver phase. A similar response would arise from a constant pulse phase as a function of the site's offset. (c) Fourier coefficients $\{A_k\}$ derived from Eq. [1], describing the SSFP response in (b); notice their rapid drop with increasing $\alpha$. (d) $\beta$-coefficients derived from a least-square solution of $\mathbf{L} \cdot \boldsymbol{\beta} = \mathbf{C}$, needed to recapitulate the illustrated filter centered at zero. (e) Actual frequency response resulting from applying the $\{\beta_m\}$-coefficients on M=50 PI-SSFP experiments upon using NB=M/2, evidencing a deteriorating resolution with increasing flip-angle.

(Fig. 2c), which increase in intensity but decay in *k*-span for increasing flip-angles. The ensuing loss in spectral information is also reflected in the {$\beta_m$} set derived from solving Eq. [5a], which manages to reconstruct the proposed frequency response with a few central *m*-values when α is small, but calls for large, widely oscillating contributions as most {$A_k$} become zero for large flip angles (Fig. 2d). This reflects an ill-conditioning of the aforementioned ***L***-matrix, and ends up leading to peak shapes that only for smaller α's, reflect the narrow filter that was originally designed (Fig. 2e). *In other words, in this PI-SSFP-based approach to high-resolution NMR, it is not only signal intensity but also spectral resolution that will be controlled by the excitation pulses.*

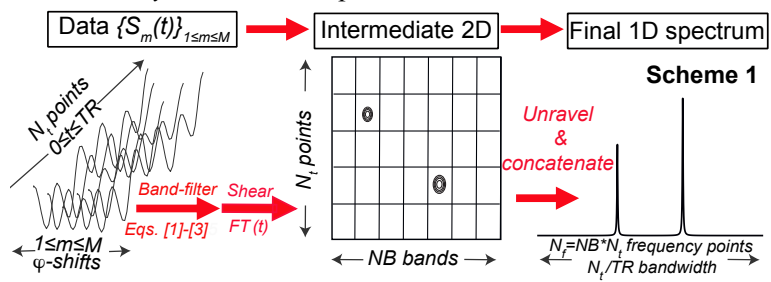

Scheme 1

The aforementioned $\beta$-coefficients can be used recursively to interrogate other contributions $I(f)$ even if $f \neq 0$, and thereby to introduce spectral resolution within the $\pm 1/2TR$ interval. In general *NB* spectral bands will thus become resolvable within this interval; assuming α has been chosen small enough to provide enough $A_k \neq 0$ coefficients, NB will then be dictated by the number *M* of phase increments used in the experiment; in our processing pipeline, we usually set *NB*=*M*/2. This $\beta$-based treatment, however, cannot address SSFP's folding problem; therefore, peaks separated by multiples of 1/*TR* (usually tens or hundreds Hz) will end up falling on the same $-\frac{NB}{2} \leq j < \frac{NB}{2} - 1$ band, and their precise resonance positions will remain unknown. As explained in Ref. 22 it is possible to endow the SSFP signal in Eq. [1] with a suitably large window that unfolds this information by sampling not just $S_m(0)$, but numerous $N_t$ points within each $0 \leq t \leq TR$ period. Performing discrete FTs on these short $S_m(t)$ FIDs will yield spectra with an overall bandwidth of $\approx 2\pi N_t/TR$, with each data point *p* separated by a frequency increment $\approx 2\pi/TR$ (in angular frequency units). As detailed in Scheme 1, adding onto this FT the phase-incremented filtering procedure described in Figure 2, can then dissect each of these spectral elements into NB finer bands. Although conceptually simple this reconstruction requires addressing a number of subtle but important issues associated with spectrometer deadtimes and band-dependent offsets, which if left unaddressed lead to spectral artifacts and phase distortions in the resulting peaks. With all these problems being deterministically addressed (Ref. 22), Figure 3 compares the performance of the resulting PI-SSFP approach vs FT-NMR $^{13}$C results, for a 5 mM sucrose sample in D$_2$O. For simplicity the Figure, as all data presented in this paper, centers on nuclear Overhauser effect (NOE) enhanced, $^1$H-decoupled solution-state acquisitions.[4,5] As can be appreciated the resolution and $SNR_t$ of both methods are comparable, raising the question of what would then be the advantage of having an SSFP-based approach that, while capable and based on its own processing pipeline, performs similarly as FT-based methods. We turn next to address this question, by describing under which conditions will PI-SSFP exceed FT-NMR's $SNR_t$ performance.

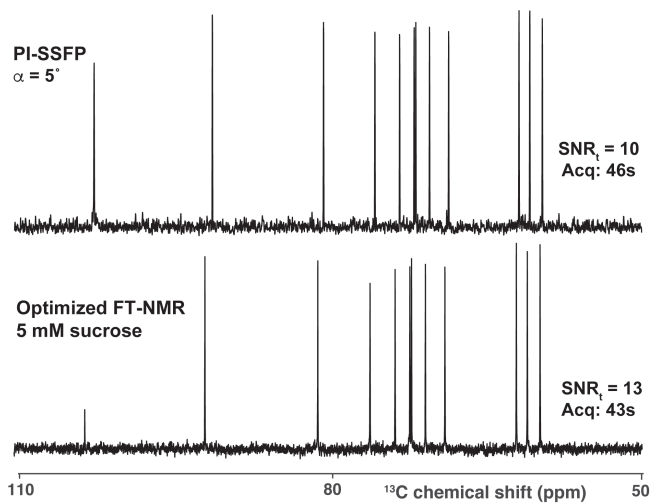

**Figure 3.** {$^1$H}$^{13}$C NMR spectra of 5 mM sucrose in D$_2$O recorded (as throughout this paper) at 14.1T using the indicated overall acquisition times. FT-NMR data was recorded using ≈50° excitation pulses, 0.6 sec acquisition times and no recovery delay (Ernst angle acquisition conditions for a T$_1$≈1.5 sec). PI-SSFP used NS=200 scans, TR= 30 ms and M=12. Shown for each experiment is the SNR$_t$ for the strongest peak in the spectrum vs noise from the (peakless) 110-140 ppm range.

**2.2 On the demands needed by PI-SSFP to overcome FT-NMR's $SNR_t$ at a given spectral resolution.** The reason why the prototypical PI-SSFP spectrum in Figure 3 does not evidence $SNR_t$ gains over its FT-NMR counterpart, derives from the relatively small flip-angle used in this acquisition. These small α-values were dictated by our search for a stable solution of the $\mathbf{L \cdot \beta \approx C}$ system of equations, needed in turn by our desire to obtain spectral linewidths in the 1-2 Hz range. On the other hand, as illustrated in Figure 1, the overall $SNR_t$ averaged over the $\pm 1/2TR$ interval for cases such as this one, where $T_1$ and $T_2$ times are relatively similar and peaks would be more-or-less randomly spread, would benefit substantially if larger flip angles were to be employed. To explore further the interplay between sensitivity and resolution in SSFP experiments, Figure 4 assumes for simplicity that resonances are uniformly distributed over $\pm 1/2TR$ (resonances outside such interval would anyhow fold over into it), and compares the averaged signal intensity afforded by SSFP acquisitions carried out as a function of flip angle, vs Ernst-angle-optimized FT-NMR. For all cases a prototypical 2 Hz resolution was assumed; this will dictate the minimal number of bands NB needed in the PI-SSFP acquisitions (disregarding for the moment the aforementioned α-effect on spectral resolution), and the duration of the FID (assumed equal to the recycle delay) for the FT-NMR case. As in our experience neither experiments nor simulations show a dependence of PI-SSFP's $SNR_t$ on $TR$ (shorter $TR$ means more phase increments are needed to finely cover the $\pm 1/TR$ bandwidth but also more scans can be packed per unit time) we also assumed that $TR$s matched –when multiplied by the number of scans and of phase increments– the overall acquisition duration of the FT-NMR FID. Given that spectral widths can also be chosen arbitrarily in both experiments by controlling the dwell times, this in turn allowed us to equalize both the overall durations of the PI-SSFP and FT NMR time-domains, and the bandwidths of the acquisitions; as a result, the noise that would affect both experiments would also be equalized. On the basis of these assumptions, it is solely the overall transverse magnetizations elicited from the spins, that will define the relative $SNR_t$ performances of the FT-NMR and PI-SSFP experiments. Figure 4 plots the relative merits that, under these assumptions, the two experiments will exhibit for $T_1$ and $T_2$ values often encountered in analytical $^{13}C$ NMR, as a function of the SSFP flip-angle employed. These plots show that in most cases, but particularly for the longer $T_2$ values favoring SSFP's $T_1=T_2$ maximum signal intensity conditions, the SSFP experiment can exceed FT-NMR's

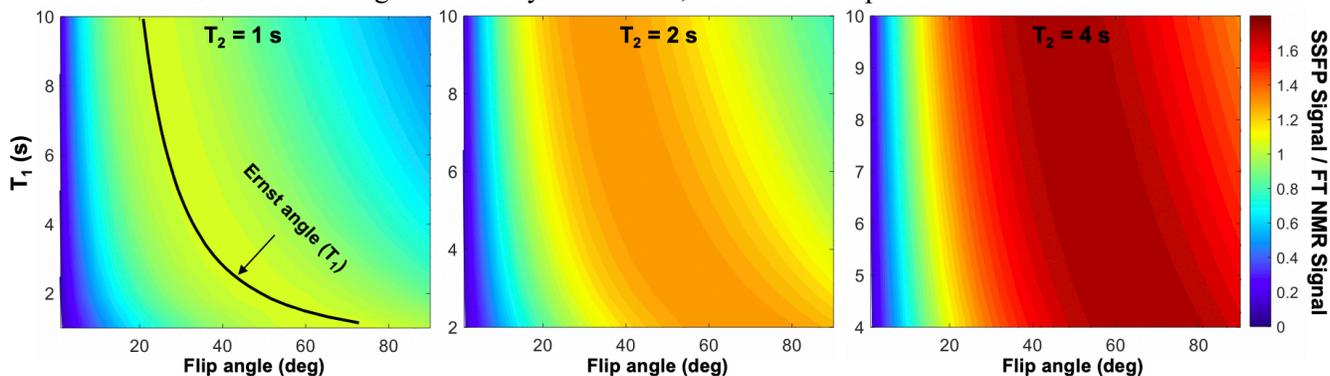

**Figure 4.** Ratio between weighted SSFP signal intensities integrated over the $\pm 1/2TR$ interval vs frequency-independent FT-NMR signal intensities as a function of the SSFP flip angle α, for a range of $T_1$, $T_2$ values. Shown for completion is the Ernst angle used for various $T_1$s (left), assuming in all cases a 2 Hz spectral resolution (see text for further details).

sensitivity –even when considering the SSFP "dark" bands. However, to achieve such superior SNR$_t$, relatively large (≥15˚) flip angles will have to be used. *And hence the dichotomy of the PI-SSFP approach as described in Paragraph 2.1: To deliver high resolution the aforementioned **L** matrix needs to be well conditioned and for this the SSFP acquisition needs to be performed using relatively small (≈5-10˚) flip angles; on the other hand, in order to maximize its sensitivity potential, SSFP data should be collected using relatively large flip angles.*

**2.3 Achieving PI-SSFP's SNR$_t$ potential: Increasing flip-angles without compromising spectral resolution.** In order to address these conflicting demands and realize PI-SSFP's SNR$_t$ potential without compromising on the approach's spectral resolution, we revisit the PI-SSFP experiment, and discuss a processing alternative that departs from the one which led to the spectrum in Figure 3. Still, in the same way as the approach used in Figure 3 had to solve an $\boldsymbol{L} \cdot \boldsymbol{\beta} = \boldsymbol{C}$ linear system of equations –and in analogy with the linear $A \cdot x = b$ equation underlying FT-NMR, where $A$ is an inverse FT matrix, $x$ the spectrum being sought and $b$ the collected FID– also the new approach to be here discussed will require solving a system of linear equations. This is because in PI-SSFP, as in FT-NMR, a linear relation links the data being collected, with the spectrum being sought. In the FT-NMR case, inverting the system of equations linking the collected FID with the sought spectrum can be done readily, thanks to the ideal conditioning of the discrete FT matrix. In PI-SSFP experiments collected as a function of phase increment $m$ and acquisition time $t$ there will also be a linear relation between the data and the high-resolution NMR spectrum. The question is what is the set of equations relating the two, and how can a stable inversion of these equations be performed, even when using the large flip angles that maximize SSFP's SNR$_t$.

To derive the linear relation in question, we rewrite the short array of FIDs collected in PI-SSFP experiments as a function of $0 \leq m \leq M-1$ phase increments (the $S_m(t)$ in Scheme 1), as a frequency-dependent construct $\boldsymbol{S_f}(t, m)$. For this we start with the t=0 expression in Eq. [1], and describe the full array of collected FIDs as

$$S_m(t) = \sum_f S_m(0) \exp(-i2\pi f t) = \sum_f I(f) \exp(-i2\pi f t) \cdot \sum_{k=-\infty}^{\infty} A_k \exp(ik\varphi_m) \cdot \exp(ik\, 2\pi f\, TR) \quad [6]$$

where the final spectrum being sought is given by the sum of all $I(f)$ amplitudes, $\boldsymbol{I} = \sum_f I(f)$. Given that the times $t$ within each FID are discretely sampled over $N_t$ equally-spaced instants, and assuming as is customary in NMR that the frequencies $f$ will be discretized over $N_f$ different values, it is possible to rewrite Eq. [6] in matrix form as

$$\underbrace{\boldsymbol{S_f}(t,m)}_{N_t \times M} = \underbrace{\boldsymbol{F}(t,f)}_{N_t \times N_f} \cdot \underbrace{\boldsymbol{I}(f)}_{N_f \times N_f} \cdot \underbrace{\boldsymbol{D}(f,m)}_{N_f \times M}. \quad [7]$$

In this Equation –which for completion shows the dimensions (rows×columns) of each of the constructs– $\boldsymbol{S_f}(t,m)$ still represents the $S_m(t)$ set of PI-SSFP FIDs, stressing now that they will be influenced by the frequency-domain peak intensities as well. $\boldsymbol{F}(t,f) = \exp(-i2\pi f t)$ is a Fourier function, discretized into a matrix among the $N_t$-values of time $0 < t < TR$ that were sampled and among the $N_f$ discrete frequencies $\{f_l\}_{0 \leq l \leq N_f - 1}$ over which the spectrum will be reconstructed. $\boldsymbol{I}(f) = \sum_l I(f_l)$ is a square $N_f \times N_f$ matrix whose only non-zero elements lie along the diagonal, and contain the high-resolution spectral information being sought, as given by an intensity $I$ for each frequency $f_l$ (i.e., $I$ is also as $N_f \times N_f$ matrix). Finally, $\boldsymbol{D}(f,m) = \sum_k A_k \exp(i2\pi m \cdot k/M) \cdot \exp(i2\pi k f TR)$ is the SSFP matrix in Eq. [1], with $f$ once again discretized over $N_f$ elements and $m$ denoting each phase-incremented acquisition.

As follows from these equations, and as further elaborated in the Supporting Information, it can be shown from Eq. [7] that for each frequency $f_l$ being interrogated, there will be a linear relationship linking the PI-SSFP signals measured, and the spectral intensity $I(f)$ being sought: $\boldsymbol{S_l}(t,m) = I(f_l) \cdot (\boldsymbol{F} \cdot \boldsymbol{D})_l$, where $(\boldsymbol{F} \cdot \boldsymbol{D})_l = \boldsymbol{G_l}$ is an $N_t \times M$ matrix resulting from calculating the $\boldsymbol{F} \cdot \boldsymbol{I} \cdot \boldsymbol{D}$ product, assuming that in $\boldsymbol{I}$ only a single line at frequency $f_l$ was present. Expanding then each $\boldsymbol{S_l}$, $\boldsymbol{G_l}$ matrix into single vectors of length $N_t M$, we end up having $\boldsymbol{S_l} = I(f_l) \cdot \boldsymbol{G_l}$, where both $\boldsymbol{S_l}$ and $\boldsymbol{G_l}$ are vectors with $N_t M$ elements. Repeating this argument for each frequency $\{f_l\}_{0 \leq l \leq N_f - 1}$, will transform $\boldsymbol{G}$ into a "supermatrix" $\boldsymbol{\Xi}$ of dimensions $N_t M \times N_f$,[a] that is related to the spectral vector as

$$\boldsymbol{S_l} = \boldsymbol{\Xi} \cdot \boldsymbol{I}(f_l). \qquad [8]$$

Here $\boldsymbol{I}(f_l)$ is an $N_f \times 1$ vector whose elements are zero for all frequencies except $f_l$, where it then takes the value $I(f_l)$. Hence, the sum of these vectors for all $f_l$, $\boldsymbol{\mathcal{I}}(f) = \sum_{l=0}^{N_f - 1} \boldsymbol{\mathcal{I}}_l(f)$, is the spectrum being sought. This sum can also be applied directly onto Eq. [8], leading to

$$\sum_{l=0}^{N_f - 1} \boldsymbol{S_l} = \boldsymbol{\Sigma} = \boldsymbol{\Xi} \cdot \sum_{l=0}^{N_f - 1} \boldsymbol{I}(f_l) = \boldsymbol{\Xi} \cdot \boldsymbol{\mathcal{I}}(f). \qquad [9]$$

The $\boldsymbol{\Xi} \cdot \boldsymbol{\mathcal{I}}(f) = \boldsymbol{\Sigma}$ form in Eq. [9] highlights the linear, $Ax=b$–type relation linking the measured PI-SSFP information in $\boldsymbol{\Sigma}$, with the spectrum residing in $\boldsymbol{\mathcal{I}}(f)$ via a transform matrix $\boldsymbol{\Xi}$. As done above for the $\boldsymbol{\beta}$-coefficients, one could in principle solve this problem in a frequency-by-frequency basis, via a least-square or pseudo-inverse solution. However, because of the considerations discussed above in connection to Figure 2, the $\boldsymbol{G_l}$ matrices will be ill-posed for inversion when the SSFP data are collected utilizing large flip-angle pulses. To solve this complication, we propose aiding the process of resolving the $\boldsymbol{\Xi} \cdot \boldsymbol{\mathcal{I}}(f) = \boldsymbol{\Sigma}$ equation with the help of regularization. In particular, we replace the solutions derived from the above-mentioned $\boldsymbol{\beta}$-coefficients, with a least absolute shrinkage and selection operator (LASSO) regression analysis,[26-28] which is particularly efficient when the solution being sought is relatively sparse –as will be the case when dealing with a high resolution NMR spectrum.[29,30] In this case, LASSO will search for the $\boldsymbol{\mathcal{I}}(f)$ "spectrum" that minimizes

$$\boldsymbol{\mathcal{L}} = \min_{\boldsymbol{\mathcal{I}} \in \mathbb{C}^{N_f}} \left\{ \frac{1}{2} \|\boldsymbol{\Xi} \cdot \boldsymbol{\mathcal{I}} - \boldsymbol{\Sigma}\|_2^2 + \lambda \|\boldsymbol{\mathcal{I}}\|_1 \right\}, \qquad [10]$$

where $\| \|_{1,2}$ stand for the first and second norm respectively, and $\lambda > 0$ is a regularization parameter promoting a solution that conforms to the general sparsity of high-resolution NMR spectra.[b] In the present study, the FISTA algorithm, which is an implementation known to solve LASSO problems efficiently, was adopted for calculating the $\boldsymbol{\mathcal{I}}(f)$ spectra.

Figure 5 presents simulations incorporating fixed noise levels and PI-SSFP data collected with a variety of flip angles, comparing the processing capabilities of the $\boldsymbol{\beta}$-filter-based proposition, with the L1 (first-norm) regularized LASSO reconstruction described in Eq. [10]. For each flip angle the signals being subject to these two processing pipelines are the same, and in all cases consisted of four peaks of relative intensities 1:0.4:0.1:0.7, affected by identical, α-independent levels of time-domain noise. These examples illustrate how the LASSO approach can deal with PI-SSFP data recorded for larger flip angles, retaining peak linewidths constant and improving the peak's intensities as the signal arising from increased flip-angles becomes larger. This is particularly evident for the peak with relative intensity of 0.1 placed at 1.0 kHz, which clearly emerges from the noise at higher αs. By contrast, PI-SSFP data processed using the $\beta$-based coefficients leads to broadened peaks when processing the larger-flip-angle FIDs, and thereby to minor decreases in the SNR despite the larger signals emitted by the spins at larger αs.

---

[a] This transitioning of matrices into vectors that are then lined up into new "supermatrices", is akin the transformation of spin operators (matrices in Hilbert space) into vectors, that are then subject to the action of "supermatrices" like Redfield's relaxation superoperator.

[b] This adaptation of a forward-fitting model to the PI-SSFP acquisition seeking to minimize the spectral L1 norm, has parallels with soft thresholding and maximum entropy approaches used in NMR and MRI.[29-31] While in the MRI case a sparsifying operation (e.g., a wavelet transformation) is needed to make the solution being sought sparse, NMR spectra of the kind being here considered do not need this extra step. See Supporting Information for further details.

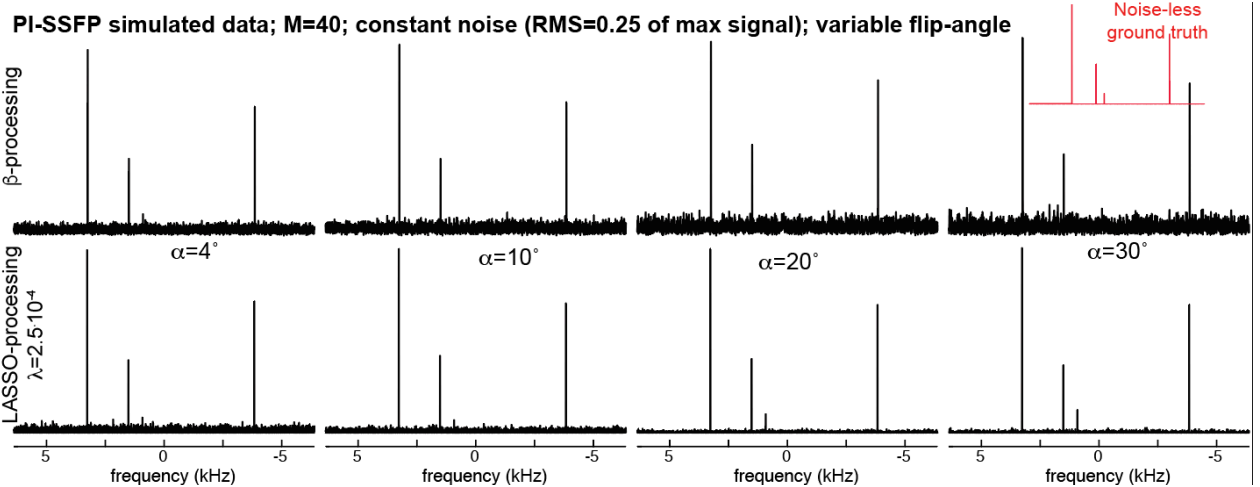

**Figure 5.** "NMR spectra" reconstructed for four uncoupled sites with relative intensities 1:0.4:0.1:0.7 positioned at 3.5, 1.6, 1.0 and -4.05 kHz, assumed subject to PI-SSFP "experiments" with different flip-angles α. Additional common parameters of all "experiments" included TR=20 ms, M=40 phase increments, 13.5 kHz sampling rates, 2 Hz digital resolutions, instantaneous (δ) pulses, identical relaxation times $T_1$ = 5 s, $T_2$= 2 s, and a constant (α-independent) noise level created by a random number generator with an RMS amplitude amounting to 25 % of the maximum single-site longitudinal magnetization (the peak at 3,5 kHz). Notice the slight SNR drop that data "acquired" for increasing α shows when processed based on the β-coefficients, despite the increase signals emitted as flip-angles increase (Figure 1). We ascribe this to a broadening of the β-derived filter functions, which is absent in the regularized version that keeps all peaks at similar half widths regardless of the flip angle used in the PI-SSFP acquisition. No such penalty affects the new, regularized reconstruction introduced in this work.

The fact that the LASSO-based processing decouples to a large extent peak line widths from flip-angles, raises in turn the issue of what will be the spectral resolution achievable by PI-SSFP data processed by this pipeline. To solve a problem like the one in Eq. [9] solely on the basis of a least-square fit, the number of rows in $\Xi$ should exceed the number of columns in the matrix; in other words, $N_t \cdot M \geq N_f$. Given that for a given *TR* the spectral bandwidth will be 1/Δt, where $\Delta t \approx TR/N_t$, the minimal frequency resolution that a least-square approach should be able to resolve will be $\Delta f \geq \frac{1}{TR \cdot M}$. When implementing the β-based reconstruction our experience was that quality spectra could be obtained upon setting $\Delta f = \frac{2}{TR \cdot M}$, provided that relatively small (≤10°) flip-angles were employed.[22] By contrast, and, thanks to the introduction of the regularization term in Eq. [10], we find that the $N_f = N_t \cdot M$ condition is often well tolerated even when using α ≈ 20-25°; for the results presented below, spectral resolution was usually set like that upon reconstruction: $\Delta f = \frac{1}{TR \cdot M}$. In principle, however, $N_f > N_t \cdot M$ also works in noiseless conditions. Notice as well that while all the elements involved in this reconstruction are complex, peaks in the ensuing $\mathcal{I}(f)$ spectrum will have no dispersive components. Hence, and although imaginary parts of these spectra are usually close to zero, the data below are plotted in magnitude mode.

## 3. Experimental

All samples investigated in this study were purchased from Sigma/Aldrich and used as received. NMR experiments were performed on a Bruker 600 MHz spectrometer using an AVIII HD console running Topspin 3.2, equipped with a TCI Prodigy® probe. The SSFP pulse sequence was written on the basis of two nested loops, whereby a train of M $\alpha_{\varphi_m}$-FID acquisition sets, possessing flip angles α and having their RF phases serially incremented by $\{\varphi_m = (m-1) \cdot \Delta\varphi\}_{1 \leq m \leq M}$, were looped M times while incrementing their Δφ's as $\Delta\varphi_k = (k-1) \cdot \frac{360°}{M}$. The receiver mode was set to "user defined" in the experiments to minimize complications arising from the machine's digital filtering, and transmitted powers were reduced to 10 W in order to have a better accuracy in the length of the flip-angles –particularly for small αs. As the interpulse time TR was only a few milliseconds, this was repeated ceaselessly NS times for the sake of signal averaging, and the ensuing signals were coadded. Given the minor changes in the phase shifts Δφ upon going from experiment k to experiment k+1 coupled to the small angles α used –leading to changes that happen with a relatively high adiabaticity throughout the M acquisitions– no dummy scans were used (although it remains to be seen how closely the steady state was then kept upon changing $\Delta\varphi_k$). Care was taken to minimize the number of points that were lost due to pulse width and receiver deadtime effects. This was done by using relatively large (40-200 kHz) receiver bandwidths, leading to short (≤20 μs) DE deadtimes; these choices did not incur in any penalties SNR-wise. All experiments used continuous GARP-based heteronuclear ¹H decoupling, leading when applicable to NOEs. When reported, relaxation times $T_1$ where measured using an inversion-recovery sequence and fitted using Topspin's relaxation toolbox. SNR was in all cases calculated as the maximum of the signal divided by the standard deviation of the noise; reported $SNR_t$ correspond to these SNR values after division by $\sqrt{total\_acquisition\_time}$. SSFP sequences were simulated and processed using Matlab-based codes, taking into account $T_1$ and $T_2$ relaxation but devoid of couplings effects.

## 4. Results

Figure 6 presents an illustrative set of results, collected on a caffeine sample. These results include an FT-NMR experiment collected in ca. 180 sec under optimized Ernst-angle conditions (top trace), and a PI-SSFP acquisition collected in a similar time using 20° flip-angle pulses and 12 phase increments. This acquisition was in turn processed using the β-based approach introduced in Ref. 22 / Figure 3 (center trace), and the new regularized reconstruction summarized in Eq. [8] (bottom trace). Highlighted on the colored insets, are the line shape improvements brought about by the new processing alternative. Also shown in the Figure is ancillary information including the peak assignment (center), the $T_1$ times measured for each site (top), and the ratio between the $SNR_t$ obtained for each peak in the FT-NMR and in the regularized PI-SSFP spectra (bottom). The actual $SNR_t$ enhancements are not important *per se* in absolute terms, as noise and sensitivity in regularized reconstructions is a complex subject requiring statistical analyses for elucidation. Still, notice how when dealing with non-protonated sites with relatively long $T_1$s, the PI-SSFP sensitivity advantage is clearly seen. Notice as well how a departure of the $T_2/T_1 \approx 1$ ratio matters in these sensitivity gains: for instance, $^{13}$Cs that are bonded to multiple $^{14}$N nuclei and have long $T_1$s but for which $T_2/T_1 \ll 1$, exhibit less impressive sensitivity gains –as reported in other SSFP studies[32]– than other quaternary carbon counterparts.

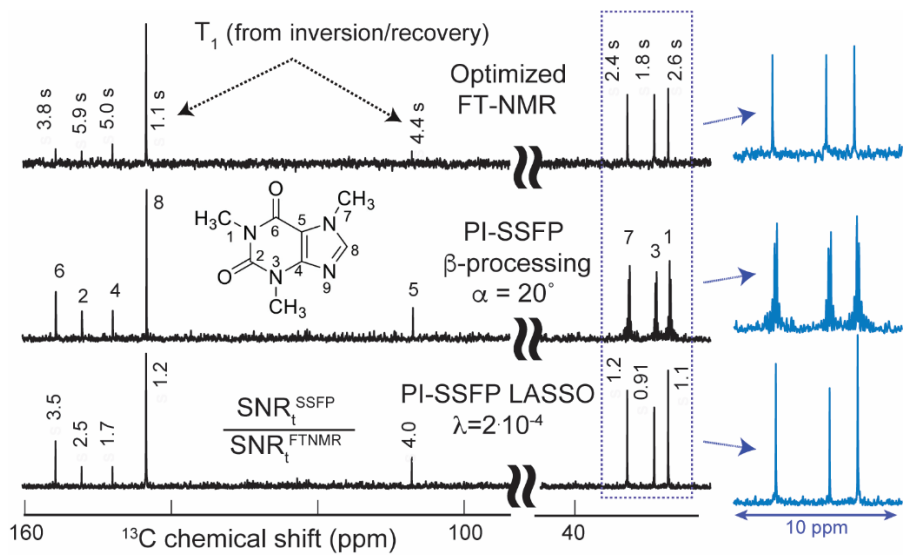

**Figure 6.** {¹H}¹³C NMR spectra of 50 mM caffeine in D₂O recorded using optimized FT-NMR (50° excitation, 0.65 s acquisition, no extra delays, 2 Hz line broadening) and PI-SSFP (TR=30 ms, M=12, α=20°) experiments, with the latter processed as shown in each row. Shown as well are $T_1$ values measured for individual peaks (top spectrum), site assignments (center spectrum), and peak-by-peak ratios between the $SNR_t$s of the LASSO-processed PI-SSFP and the FT-NMR (bottom spectrum). Notice how slowly relaxing sites benefit the most $SNR_t$-wise from PI-SSFP, and how minor artifacts in PI-SSFP data processed based on the β-coefficients disappear in the LASSO pipeline. All data were collected in ca. 180 sec. Empty spectral regions were cropped away for clarity.

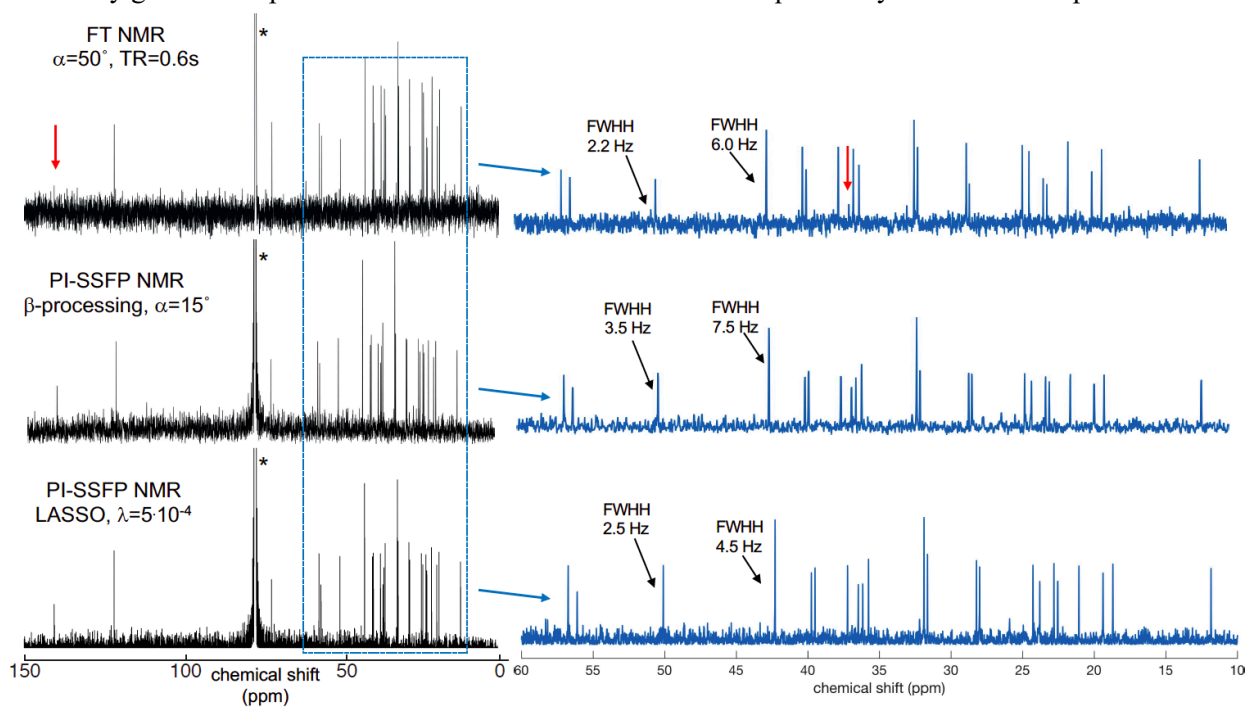

**Figure 7.** 1D {$^1$H}$^{13}$C FT-NMR spectrum (top) and PI-SSFP NMR spectra processed via the previously proposed (center) and the newly regularized (bottom) approaches, on the same 7 mM cholesterol solution in CDCl$_3$ (cropped peak, indicated by an asterisk). Both sets were collected in 40 s; traces in blue show zoom-ins to the 10-60 ppm regions. Listed for each trace are main experimental acquisition conditions; the PI-SSFP data involved M=12 phase shifts with a TR=30ms. Indicated are the full-widths at half-height (FWHH) of representative peaks. Red arrows highlight quaternary carbons $C_1$ and $C_{12}$ at 141 and 36.5 ppm, possessing longer $T_1$s that negatively bias sensitivity in conventional FT-NMR but not in the new PI-SSFP experiment. All remaining visible peaks have shorter ≤1s $T_1$s.

Figures 7 and 8 highlight the capability of the new processing procedure to deliver excellent peak shapes over arbitrary bandwidths, while exhibiting some clear $SNR_t$ advantages over FT-NMR. Notice in particular how PI-SSFP's sensitivity will match that of FT-NMR for most of the peaks, but make sites with longer $T_1$s –whose signals are sometimes buried in the FT-NMR spectral noise– appear with good sensitivity within similar acquisition times. Notice as well that while relying on relatively large flip angles, an improved spectral resolution characterizes the LASSO reconstructions over the β-based counterparts. The advantages seem particularly large in the case of the non-protonated amines in Figure 8, which since devoid of $^1$H NOEs and given $^{15}$N's low natural abundance, present a significant challenge to FT-NMR. The long $T_1$s of these non-protonated sites further compounds this problem; although these were not evaluated for all the assayed compounds, $T_1$ was found to be on the order of 40 s for pyridine's $^{15}$N, and on the basis of this value the excitation pulses used in FT-NMR acquisition were set. Judging by the observed peak intensities, it appears that for some of the remaining compounds examined the $^{15}$N $T_1$s were even longer; in all these cases, the advantages of PI-SSFP were most evident.

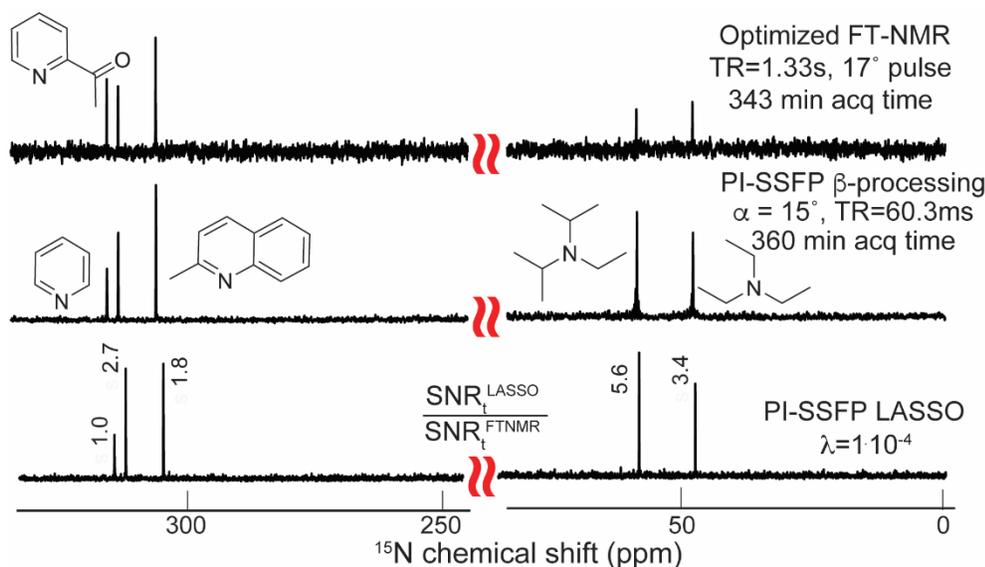

**Figure 8.** $\{^1H\}^{15}$N NMR spectra of equal volumes (50 μL each) of the five indicated nitrogenated bases, dissolved in 250 μL of d$_6$-DMSO. The FT-NMR data were processed with a 2 Hz line broadening, and its line widths ranged between 2.0 and 2.3 Hz. The PI-SSFP β-processed linewidths ranged between 4.0 and 4.6 Hz, whereas the LASSO-derived line widths were ≈1.9-2.4 Hz. Data were collected in nearly the same acquisition times, leading to the SNR$_t$ ratios between the PI-SSFP LASSO and FT-NMR data indicated next to each peak (the ratios for the β-processed peaks were similar). Empty spectral regions were cropped away for clarity.

## 5. Discussion and Conclusions

The present study was motivated by a search for endowing the recently-proposed PI-SSFP approach to high-resolution NMR, with the sensitivity advantages known to characterize SSFP for $T_1 \approx T_2$ conditions. These sensitivity advantages will come into play foremost when seeking resolution and having to deal with sites with long $T_1$s: FT-NMR resolution considerations will then demand the use of relatively long acquisition times TR, while Ernst-angle constraints will then require small flip-angles $\alpha_E = \cos^{-1}(e^{-TR/T_1})$ to maximize spectral SNR$_t$. By contrast no such long-TR demands will affect PI-SSFP, whose resolution is given by the flip-angle and the number of employed phase increments. However, to make full use of the ensuing competitiveness, one still needs to have the option of collecting the PI-SSFP experiments over a range of flip angles. In particular, when $T_1 \approx T_2$, optimal SSFP SNR$_t$ conditions will demand working with flip-angles in the ≥40° range; according to our previously-proposed implementation, this would bring about an ill-conditioning of the data processing, and lead to a broadening in the point-spread functions characterizing the peaks. By casting the problem in a different, *Ax=b* fashion susceptible to a least-square regularized inversion, this study lifts such limitation. Regularization then enabled us to work at high flip-angles without compromising the spectral resolution; notice, however, that whenever dealing with very long $T_1$s –including in several of the sites in the compounds here assayed– the $T_2 < T_1$ condition will still require relatively small flip angles for an optimal SNR$_t$.

The demonstration that PI-SSFP signals and 1D NMR spectra are linearly related to one another (Eqs. [8], [9]), endows the PI-SSFP experiment with a number of processing alternatives. The present study adopted one of them, based on an L1-regularized procedure. Other regularization options were also tested, but LASSO's FISTA-based implementation provided the most satisfying results in terms of line shapes, speed and robustness. Unfortunately, the use of regularized least-square procedures to reconstruct a spectrum, is also associated to drawbacks and challenges. In terms of drawbacks, the most evident one is paid in terms of computation: whereas processing an 8 kHz spectrum with a 1.3 Hz resolution takes ca. 20 sec on a 16-core i7 desktop computer when implemented by the iterative LASSO approach, the same spectrum was processed in 0.5 sec by the non-iterative β-coefficient formalism (which then lead to ≥3 Hz spectral resolutions). A second drawback, associated more with the appearance one expects from an NMR spectrum than with the essence of the information, relates to the positive-only nature of the spectrum that arises from the new iterative fit; spectral intensities are naturally expected to be positive only, but the non-regularized, least-squares nature of the β-coefficient processing, provides both positive and negative values as a result of its fit. Besides these technical details, more complex features are associated with the LASSO approach. One of them is difficulty in quantifying SNR: once a peak exceeds the noise level, the value chosen for the regularization factor can dominate the appearance of a spectrum. A complex literature exists about this fact, including guides on how optimally set the regularization parameter, and how to estimate sensitivity in the sense of being routinely able to distinguish a genuine peak out of random noise via statistical analyses.[26-28] We are still in the process of investigating these matters, in the search for what the optimal estimator should be for the present NMR application. This, however, does not mean the processing is not quantitative in terms of the relative heights of the peaks –in that sense, the new pipeline is very well behaved. Also establishing the ultimate resolution of the method becomes more challenging upon invoking regularization: indeed, when relying on the β-based processing, relatively straightforward relations linked the width of the PI-SSFP point-spread-function to the flip-angle and to the number of phase increments that needed to be collected in order to distinguish the various $A_k \neq 0$ coefficients. However, upon including a regularization process, spectral resolution can be substantially increased beyond the aforementioned $\Delta f = \frac{1}{TR \cdot M}$ limit. The amount of noise will then play an important role on how large can such resolution enhancement be: if the spectral data is sparse (which it usually is) and devoid of noise, as few as $M=3$ increments may suffice to unravel even complex multiline spectra; basically, as long as $MN_t$ is larger than the number of peaks, the spectrum can be recapitulated with the help of the regularization. This in turn also highlights the fact that, while we have been performing comparisons against NMR data processed by the usual discrete FT, comparisons against alternative, regularized-based processing approaches might also be relevant. These might make Ernst-angle-based acquisitions more competitive –even if this would not be able to overcome the physics-based insight from Figure 4, showing that spins in SSFP-based experiments will generate the highest signals per unit time.

Another factor that remains to be unraveled concerns how will $T_1$ and $T_2$ times affect the PI-SSFP line shapes. The influence of these times is very well understood in multi-scan FT-NMR: here $T_1$ will define the signal intensity via the Ernst angle, and $T_2$ the line width and the optimal matching weighting function to be used. The actual FT inversion relating the FID to the spectrum, however, is independent of these time constants. By contrast, $T_1$ and $T_2$ enter in the definition of the Ξ PI-SSFP transformation matrix, meaning that line shapes may become affected if these are unknown or very wrongly assumed. While we have not found significant evidence for such effects in numerical simulations, this is a feature that remains to be further investigated. Another intriguing possibility concerns what will happen if steady-states are not necessarily reached over the PI-SSFP procedure; SSFP experiments have, after all, been extensively used in hyperpolarized NMR imaging, under scenarios where polarizations are rapidly decreasing over the course of the pulsing.[33-35] We hypothesize that even under such cases, suitable processing avenues could still enable the acquisition of high resolution NMR spectra via phase-incremented, rapid-pulsing approaches, by relying on the linear relations and principles described above. Additional ways of exploiting this linearity to retrieve spectra that are not based on regularized reconstruction, can also be devised. These and other aspects of this intriguing new route to high-resolution NMR, will be discussed in upcoming studies.

**Data Availability:** All data and simulations described in the present work are available from the authors upon reasonable request.

**Acknowledgements.** We are grateful to Dr. Julia Grinshtein for assistance in this study. This work was funded by the Israel Science Foundation grant 1874/22, by the Minerva and the Helmholtz Foundations (Germany), and by the generosity of the Perlman Family Foundation. LF holds the Bertha and Isadore Gudelsky Professorial Chair.

**Author Contributions.** YZ and LF conceived the project. MS, TH, EM and LF developed methodologies and collected/processed the data. AL and YZ contributed calculations. All authors discussed the results leading to the final manuscript. LF and YZ wrote the paper.

**Ethics declarations.** The authors have no competing interests to declare.

**Supplementary Information:** A detailed description about the new physics and LASSO-based processing of the phase-incremented SSFP data in 1D decoupled NMR acquisitions.

# Supplementary Information for

## Phase-Incremented, Steady-State Solution NMR: Maximizing Spectral Sensitivity Without Compromising Resolution

Mark Shif,[1] Yuval Zur[2,*], Adonis Lupulescu,[1] Tian He,[1,3] Elton T. Montrazi,[1] and Lucio Frydman[1,*]

[1]Chemical and Biological Physics Department, Weizmann Institute, Rehovot, [2]Insightec Ltd, Tirat Carmel, Israel; [3]Chemistry Department, Zhejiang University, P. R. China

*Emails: yuvalzur50@gmail.com; lucio.frydman@weizmann.ac.il

### Introduction: Revisiting the SSFP signal

The SSFP sequence consists of an equidistant train of RF pulses with flip angle a and phase q. The time between these RF pulses is TR << T2. The time t ($0 \leq t \leq TR$) between adjacent RF pulses, is t = 0 immediately after the RF pulse and t = TR immediately prior to the next RF pulse. The signal S(t) from a spin with angular frequency $\omega = 2\pi f$ is

$$S(t) = S(0, \Phi) \exp(-i\omega t) = S(0, \Phi)\exp(-i\varphi(t)) \quad [S1]$$

$\varphi(t) = 2\pi f t$ is the phase accrual of the spin from time 0 to t. $S(0, \Phi)$ is the signal at t = 0 which depends on the phase accrual $\Phi$ (from prior TR's) from t = 0 to t = TR:

$$\Phi = 2\pi f\, TR = 2\pi \cdot n + \phi \quad [S2]$$

where n is an integer and $-\pi \leq \phi \leq \pi$. After a few RF pulses a steady-state is established, such that $S(0, \Phi)$ is the same for all subsequent TR's. $S(0, \Phi)$ is a function of the spectral intensity $I$, RF flip angle $\alpha$, RF phase $\theta$, $T_1$, $T_2$ and TR. As described in Ref. (1):

$$\mathbf{S}(0, \Phi) = \exp(i\psi) \cdot I(\Phi) \cdot \frac{a \exp(i\Phi) + b}{c \cos(\Phi) + d} \quad [S3]$$

where $\psi = \theta + \frac{\pi}{2}$ is the phase of the spin at t = 0 and $I(\Phi)$ is the spectral intensity of a site that precessed a phase $\Phi$ –i.e., at a frequency $\omega$ for a known time $TR$. a, b, c and d are functions of $T_1$, $T_2$, $\alpha$ and TR, but are independent of $\Phi$. $S(0, \Phi)$ is periodic in $\Phi$ modulus $2\pi$, such that $\Phi$ in Eq. [S3] can be replaced with the expression in Eq. [S2].

### On the linearity between NMR's high-resolution spectrum and the PI-SSFP set of signals

Our purpose is to find the high-resolution NMR spectrum $I(f)$, for all possible frequencies $f$. For the short TR's required by the SSFP conditions, a Fourier Transform (FT) of the $S(t)$ FID will not suffice for this; for instance, for TR ~10 ms, 1/TR will be 100 Hz –ca. two orders of magnitude larger than the kind of linewidths high resolution NMR is seeking. Furthermore, the SSFP conditions will generate periodic distortions in the peak amplitudes, including so-called "dark bands", further distorting the spectrum. To solve these problems PI-SSFP acquires M sets of SSFP signals, with the excitation phases between consecutive scans shifted by $m\frac{2\pi}{M}$ radians (2). In other words, it collects a set of M independent SSFP FIDs (each of this signal averaged as need be), where

$$\Phi_m = \Phi + m\frac{2\pi}{M}, \qquad \text{where } m = -\frac{M}{2} \text{ to } \frac{M}{2} - 1 \qquad [S4a]$$

The signal $S_m(f,t)$ of the m$^{th}$ scan for a spin with frequency $f$ will then be

$$S_m(f,t) = S(0,\Phi_m)\exp(-i2\pi ft) \qquad [S4b]$$

We assume that our high-resolution spectrum $\mathfrak{I}$ consists of spectral lines at discrete frequencies $f_i$ with amplitudes $I$. So

$$\mathfrak{I}(f) = \sum_i I(f_i) \qquad [S5]$$

During each SSFP scan we sample $N_t$ points over a time TR. The spectrum $\mathfrak{I}$ is thus assumed contained within the spectral bandwidth $SW = \frac{N_t}{TR}$. For a spectral resolution $\Delta f$, the number of points $N_f$ along the frequency axis of the spectrum will thus be $N_f = \frac{SW}{\Delta f}$.

Since $\Phi = 2\pi f \cdot TR$ (Eq. [S2]), and $\Phi_m = \Phi + m\frac{2\pi}{M}$ (Eq. [S4a]), $S(0,\Phi_m)$ in Eq. [S4b] can also be written as

$$S(0,\Phi_m) = I(f) \cdot \frac{a\exp\left[i\left(2\pi f \cdot TR + m\frac{2\pi}{M}\right)\right] + b}{c\cos\left(2\pi f \cdot TR + m\frac{2\pi}{M}\right) + d} \qquad [S6]$$

where the $\exp(i\psi)$ in Eq. [S3] was absorbed into a and b. $S(0,\Phi_m)$ in Eq. [S6] can be written as a $N_f$-by-$M$ matrix, where $N_f$ is as mentioned the length of the frequency vector, and M is the number of phase-increments in the PI-SSFP scans. Given this discretization we thus write $S(0,\Phi_m)$ as $S(f,m)$, an $N_f$-by-M matrix. $I(f)$ in [S6] is a sparse vector which is non-zero at a finite number of frequencies $f_i$ (Eq. [S5]), and zero where there are no peaks.

Given Eq. [S6] we can write $S(f,m)$ as a product of a diagonal $N_f$-by-$N_f$ matrix $I$, and the $N_f$-by-$M$ matrix $D$:

$$S(0,\Phi_m) \triangleq S(f,m) = I(f) \cdot D(f,m) \qquad [S7]$$

$I(f)$ has along its diagonal the N-by-1 vector $\mathfrak{I}(f)$ from Eq. [S5]. The N-by-M matrix $D(f,m)$ is given by

$$D(f,m) = \frac{a\exp\left[i\left(2\pi f \cdot TR + m\frac{2\pi}{M}\right)\right] + b}{c\cos\left(2\pi f \cdot TR + m\frac{2\pi}{M}\right) + d} \qquad [S8]$$

As described in Eq. [4] of the main text, the measured signal of the m-shifted SSFP scan at a time t, $0 \le t \le TR$, for an ensemble of spins with many frequencies, will be

$$S_m(t) = \int S_m(f,t)\,df = \int S(f,m)\exp(-i2\pi ft)\,df \qquad [S9]$$

The discrete version of Eq. [S9] will then be the sum over the discrete $N_f$ frequency values taken by vector $f$:

$$S_m(t) = \sum_f \exp(-i2\pi ft) \cdot S(f,m) \qquad [S10]$$

Given that the time variable $t$ has also been discretized over $N_t$ values for each collected FID, we can define a Fourier matrix $F(t,f)$

$$F(t,f) = \exp(-i2\pi ft) \qquad [S11]$$

which is a $N_t$-by-$N_f$ matrix. From Eqs. [S7] and [S10], the measured signal **S** of all M SSFP-shifted scans and all times t ($0 \leq t \leq TR$) can therefore be described as:
$$\mathbf{S} = \mathbf{F}(t,f) \cdot \mathbf{I}(f) \cdot \mathbf{D}(f,m) \qquad [S12]$$
Eq. [S12] is a key Equation since it expresses the linear relation that ties the PI-SSFP signals **S** that were measured, and the $\mathbf{I}(f)$ spectrum being sought. It also implies that the achievable spectral resolution of $\mathbf{I}(f)$ in PI-SSFP, i.e. the span of f and how many experiments *M* and samplings $N_t$ will be required to resolve each $N_f$ element, will depend on the accuracy of the solution method that can be used to calculate $\mathbf{I}(f)$ upon solving this linear system of equations.

**Measured signal of a single spectral line at $f_k$**

To solve Eq. [S12], we consider a single spectral line at frequency $f_k$; i.e. we assume the k$^{th}$ element of the frequency vector *f* has an amplitude $I_k$, and amplitudes are zero at all other frequencies. In that case the matrix $\mathbf{I} \cdot \mathbf{D}$ in Eq. [S13] has only one non-zero row (of length M); this shows at row k, and within the SSFP approximation it will be given by (see Eq. [S6])

$$(\mathbf{I} \cdot \mathbf{D})_k = I(f_k) \cdot \frac{a \exp\left[i\left(2\pi f_k \cdot TR + m\frac{2\pi}{M}\right)\right] + b}{c \cos\left(2\pi f_k \cdot TR + m\frac{2\pi}{M}\right) + d} = I(f_k) \cdot \mathbf{d}_k \quad m = -\frac{M}{2} \text{ to } \frac{M}{2} - 1 \quad [S13]$$

The matrix $(\mathbf{I} \cdot \mathbf{D})_k$ k$^{th}$ row is given by Eq. [S13], and is zero elsewhere.

Denoting the k$^{th}$ column of matrix **F** in Eq. [S11] as $\mathbf{F}_k = \exp(-i2\pi f_k t)$, then the sampled signal $\mathbf{S}_k$ in Eq. [S12] becomes
$$\mathbf{S}_k = I(f_k) \cdot (\mathbf{F} \cdot \mathbf{D})_k \triangleq I(f_k) \cdot \mathbf{G}_k \qquad [S14]$$
where $\mathbf{G}_k \triangleq (\mathbf{F} \cdot \mathbf{D})_k$ is the $\mathbf{F} \cdot \mathbf{D}$ product for the case when there is only one spectral line at $f_k$. The j$^{th}$ row of $\mathbf{G}_k$ is obtained by multiplying each element of the row vector $\mathbf{d}_k$ in Eq. [S13] by the j$^{th}$ element of $\mathbf{F}_k$:
$$j^{th} \text{ row of } \mathbf{G}_k = \exp(-i2\pi f_k t_j) \cdot \mathbf{d}_k \qquad [S15]$$
where $t_j$ is the j$^{th}$ time point in TR. Both $\mathbf{S}_k$ and $\mathbf{G}_k$ in Eq. [S14] are $N_t$-by-M matrices. To simplify, we concatenate all the columns of $\mathbf{S}_k$ and $\mathbf{G}_k$ into column vectors with $N_t$*M elements:
$$\mathcal{S}_k = I(f_k) \cdot \mathcal{G}_k \qquad [S16]$$
where $\mathcal{S}_k$ and $\mathcal{G}_k$ are the vectorized vectors of matrices $\mathbf{S}_k$ and $\mathbf{G}_k$ from a single spectral line at frequency $f_k$.

**From the measured signal to spectral intensities at arbitrary frequencies**

The vector $\mathcal{G}_k$ in Eq. [S16] can be calculated for any arbitrary frequency $1 \leq f_l \leq N_f$. All the vectors $\mathcal{G}_n$ can then be arranged into a single $N_t$*M-by-$N_f$ matrix $\Xi$, which is known. Also the $I(f_k)$ in Eq. [S16] can be arranged as a $N_f$-by-1 vector $\mathfrak{I}_k$ which is zero for all the frequencies other than k, and its k$^{th}$ entry is $I(f_k)$. Using these definitions, we rewrite Eq. [S16] as:
$$\mathcal{S}_k = \Xi \cdot \mathfrak{I}_k \qquad [S17]$$
where $\Xi$ is an $N_t$M-by-$N_f$ matrix and $\mathfrak{I}_k$ is a $N_f$-by-1 vector. Eq. [S17] hold for a single spectral line at bin k. In the general case, for the many spectral lines in the frequency vector $f_k$, the total spectrum $\mathfrak{I}$ will be the sum over all $N_f$ bins:

$$\mathfrak{I}(f) = \sum_{k=1}^{N_f} \mathfrak{I}_k \qquad [S18]$$

Likewise, the $N_t$M vector of the experimentally measured matrix $\mathbf{S}$ in Eq. [S12], will be the sum of all the $\mathbf{S_k}$ in Eq. [S17] –also summed over all $N_f$ spectral bins:

$$\mathbf{\Sigma} = \sum_{k=1}^{N_f} \mathbf{S_k} = \mathbf{\Xi} \cdot \sum_{k=1}^{N_f} \mathfrak{I}_k = \mathbf{\Xi} \cdot \mathfrak{I}(f) \qquad [S19]$$

Our goal is to compute the spectrum $\mathfrak{I}(f)$ from the measured vector $\mathbf{\Sigma}$ and the known matrix $\mathbf{\Xi}$.

**Solving for $\mathfrak{I}(f)$**

Eq. [S21] is a classic $\mathbf{Ax} = \mathbf{b}$ linear system. To solve it we rely on the fact that the spectrum $\mathfrak{I}(f)$ contains a relatively small number of sharp peaks, i.e. it is sparse. Such a problem can be solved using a number of algorithms; here we adopted the LASSO approach (3), where the first norm (sum of magnitude values) of $\mathfrak{I}(f)$ is added as a regularization term. LASSO then solves the problem by computing the $\mathfrak{I}$ that minimizes the combination $\mathcal{L}$ defined as

$$\mathcal{L} = \frac{1}{2} \|\mathbf{\Xi} \cdot \mathfrak{I} - \mathbf{\Sigma}\|_2^2 + \lambda \|\mathfrak{I}\|_1 \qquad [S20]$$

where the subindexed 2 and 1 stand for the second and first norms L2 and L1, and $\lambda > 0$ is a regularization factor promoting a sparse solution, which is in line with the prior knowledge that $\mathfrak{I}$ is sparse. In the present study, we use the FISTA implementation of the LASSO approach (4), as this in known to solve problems like Eq. [S20] efficiently.

If a standard least-squares solution would be utilized to get $\mathfrak{I}$ from Eq. [S19], the matrix $\mathbf{\Xi}$ would have to be "tall"; i.e., possess more rows (data) than columns (unknowns). Since $\mathbf{\Xi}$ has $N_t$M rows and $N_f$ columns, this limits the number of frequency bins that could be solvable to $N_f \leq N_t * M$. This in turn would limit the spectral resolution to $\Delta f \geq \frac{1}{TR \cdot M}$. The assumption that the overall spectrum $\mathfrak{I}$ is sparse, on the other hand, lifts this constraint, and enables the matrix $\mathbf{\Xi}$ to be "fat"; i.e., to have more unknown frequency elements than measured data. This is not surprising, as given the regularized assistance it is the number of non-zero peaks rather than the number of frequency bins, which defines the size required to reach a given spectral resolution. This is well known from non-uniform and compressed sensing approaches widely used in multidimensional MRI (5) and NMR (6). In the former case, as in many multidimensional image compression algorithms, a wavelet-transform needs to be applied to the image before reaching a representation that is sparse. This is not necessary in already sparse multidimensional NMR acquisitions; the aforementioned procedure also adds this now to the practical benefit of 1D NMR.

**References – Supporting Information**
(1) Y. Zur, M. L. Wood and L. J. Neuringer, Motion-insensitive steady-state free precession imaging, *Mag. Res. Med* **16**, 444 – 459 (1990).
(2) He, T., Zur, Y., Montrazi, E. T., and Frydman, L. Phase-Incremented Steady-State Free Precession as an Alternate Route to High-Resolution NMR. *J. Am. Chem. Soc.* **146**, 3615–3621 (2024).
(3) Robert Tibshirani. Regression shrinkage and selection via the lasso. *Journal of the Royal Statistical Society. Series B (Methodological),* 267–288 (1996).